\documentclass{article}
\usepackage{spconf,amsmath,graphicx}
\usepackage[hyperfootnotes=false]{hyperref}
\usepackage{booktabs}
\usepackage{amssymb}
\usepackage{multirow}
\usepackage{etoolbox}
\usepackage{multicol}
\usepackage{placeins}
\hypersetup{hidelinks}
\makeatletter
\patchcmd{\@makecaption}{\vskip 10pt}
  {\ifdefstring{\@captype}{figure}{\vskip 5pt}{\vskip 10pt}}
  {}{\PackageError{TemplateCraft}{Could not set figure caption spacing}{Check the spconf caption definition.}}
\makeatother

\makeatletter
\patchcmd{\section}{-3.5ex \@plus -1ex \@minus -.2ex}{-\baselineskip}
  {}{\PackageError{TemplateCraft}{Could not set section space before}{Check the section definition.}}
\patchcmd{\section}{2.3ex \@plus.2ex}{\baselineskip}
  {}{\PackageError{TemplateCraft}{Could not set section space after}{Check the section definition.}}
\makeatother

\makeatletter
\patchcmd{\subsection}{-3.25ex\@plus -1ex \@minus -.2ex}{-3pt}
  {}{\PackageError{TemplateCraft}{Could not set subsection space before}{Check the subsection definition.}}
\patchcmd{\subsection}{1.5ex \@plus .2ex}{3pt}
  {}{\PackageError{TemplateCraft}{Could not set subsection space after}{Check the subsection definition.}}
\makeatother

\makeatletter
\patchcmd{\subsubsection}{-3.25ex\@plus -1ex \@minus -.2ex}{-4pt}
  {}{\PackageError{TemplateCraft}{Could not set subsubsection space before}{Check the subsubsection definition.}}
\patchcmd{\subsubsection}{1.5ex \@plus .2ex}{2pt}
  {}{\PackageError{TemplateCraft}{Could not set subsubsection space after}{Check the subsubsection definition.}}
\makeatother

\makeatletter
\patchcmd{\@maketitle}{\large \bf \@title}
  {\fontsize{14pt}{16.8pt}\selectfont\bfseries \@title}
  {}{\PackageError{TemplateCraft}{Could not apply 14pt title formatting}{Check the spconf title definition.}}
\makeatother

\title{TemplateCraft: Agentic Visual Template Generation}
\name{\shortstack{Hongjie Yu$^{1,2}$, Zhiyuan Fan$^{1}$, Yuzhe Zhang$^{1}$,
Jiangcun Du$^{2}$,\\[1pt]
Zhicheng Gao$^{2}$, Yuhong Zhang$^{2}$, Xiaokai Zhan$^{2}$,
Zongshi Xie$^{2,*}$}%
\thanks{\fontsize{9pt}{10.4pt}\selectfont\raggedright * Corresponding author: Zongshi Xie, \href{mailto:xiezongshi@kuaishou.com}{\mbox{xiezongshi@kuaishou.com}}.}}
\address{$^{1}$Peking University \qquad $^{2}$Kuaishou Technology}

\begin{document}
\ninept
\setlength{\abovedisplayskip}{3pt}
\setlength{\belowdisplayskip}{3pt}
\setlength{\abovedisplayshortskip}{3pt}
\setlength{\belowdisplayshortskip}{3pt}
\begingroup
\makeatletter
\long\def\@makefntext#1{\noindent#1}
\makeatother
\maketitle
\endgroup
\makeatletter
\long\def\@makefntext#1{\noindent\@makefnmark\,#1}
\makeatother

\begin{abstract}
The growing popularity of short videos has driven demand for one-click content creation. Visual
templates turn uploaded images into personalized content with preset effects, but reusable template
generation still requires substantial manual effort in asset preparation and tool orchestration. We propose
TemplateCraft, a multi-agent system that converts natural-language instructions into client-executable
templates through planning, material generation, effect-workflow generation, and protocol compilation.
Its Planner--Evaluator loop uses execution feedback for targeted rollback, while stage-level and long-term
memory support revision without parameter updates. We evaluate TemplateCraft on
TemplateBench\footnote{\fontsize{9pt}{10.4pt}\selectfont The benchmark will be released upon acceptance:
\url{https://github.com/Xiaoyu-Fish-hub/TemplateBench}.}, derived from 60 real-world templates. With the same
Qwen3-VL backbone, TemplateCraft raises image/video generation success rates from 56.7\%/30.0\% to
66.7\%/50.0\% over Planner-only (best-of-three) and improves template adherence and style consistency.
With additional evaluation and revision, it matches or exceeds a GPT-4o Planner-only baseline on selected
metrics. Persistent assets further improve cross-input style consistency.
\end{abstract}


\begingroup
\setlength{\parskip}{0pt}
\section{Introduction}
\label{sec:introduction}
\stepcounter{footnote}
\begingroup
\makeatletter
\long\def\@makefntext#1{\noindent\@makefnmark\,#1}
\makeatother
\footnotetext{\fontsize{9pt}{10.4pt}\selectfont Kwai's official video editing tool.}
\endgroup

A portrait-restyling template can be reused by many users: each user uploads a photograph, and the
template applies a predefined clothing style and scene through a fixed sequence of transformations to
produce personalized content. We refer to such a reusable workflow as a visual template. It encapsulates
the creative decisions and production operations, reducing the burden of tool selection and workflow
orchestration on users. Creating such templates,
however, still requires designers to devise effects, prepare assets, and repeatedly refine the
pipeline. New themes and styles often demand new templates, leaving the conversion of creative
ideas into deployable templates heavily dependent on manual effort.

Large language model agents provide a foundation for automating this process. Toolformer
~\cite{schick2023toolformer}, ReAct~\cite{yao2023react}, and ToolLLM
~\cite{qin2024toolllm} study tool-use learning, interleaved reasoning and acting through environment
interaction, and large-scale API learning and invocation, respectively. HuggingGPT~\cite{shen2023hugginggpt} and Visual ChatGPT
~\cite{wu2023visualchatgpt} extend tool orchestration to multimodal tasks, while ComfyGPT
~\cite{huang2025comfygpt}, ComfyUI-R1~\cite{xu2026comfyuir1}, and ComfyUI-Copilot
~\cite{xu2025comfyuicopilot} further support workflow generation. In cinematic production,
FilmAgent~\cite{xu2025filmagent} and MovieAgent~\cite{wu2025movieagent} coordinate
scriptwriting, storyboarding, and multi-shot generation through role specialization. Related studies
also address animation generation, human--AI collaboration during pre-production, and
feedback-guided editing~\cite{li2024animdirector,wang2025animagents,huang2025filmaster}.

Although these studies support content generation and workflow orchestration, they primarily target
one-off content generation and conventional video production. To the best of our knowledge,
reusable visual-template generation remains underexplored. Turning a generated
workflow into a deployable template further requires defining replaceable user inputs, binding
persistent assets, and connecting both to the appropriate generation steps. A single successful execution is
also insufficient to establish how well a workflow can be reused after its inputs are replaced. We
therefore study how to generate complete visual templates from natural-language instructions, so
that different users can execute the same workflow while preserving the intended creative effect.
This task requires a model to coordinate input constraints and dependencies across steps while
detecting outputs that fail to meet the requirements despite successful execution. For example, a portrait-restyling
result may exhibit the target aesthetic while failing to preserve the input subject. Synchromesh
~\cite{poesia2022synchromesh} reduces program-generation errors through syntactic and semantic
constraints, whereas template generation additionally requires assessing intermediate artifacts and
visual outputs and identifying where revision should occur.

To address this problem, we propose TemplateCraft, which decomposes template generation into four
stages: template planning, material generation, effect-workflow generation, and protocol
compilation. The first three stages are governed by a Planner--Evaluator loop. The Planner generates
specifications and workflows from the creative instruction, category skills, and tool constraints,
and an executor invokes the selected tools. The Evaluator diagnoses problems from execution
feedback and candidate outputs and directs the system to the earliest stage requiring revision. The
revised template is then compiled and packaged, and users replace only the designated inputs at
deployment. Inspired by Reflexion~\cite{shinn2023reflexion}, which uses feedback and experiential
memory to guide subsequent attempts, TemplateCraft employs stage memory for targeted revision of
the current task and long-term memory to store diagnostic knowledge for reuse across tasks, without
updating model parameters.

To evaluate template reuse across inputs, we construct TemplateBench from production templates in
KwaiCut\footnotemark[2]. It contains 30 image-template and 30 video-template tasks, each with a
creative instruction, scoring rules, and five test input sets. Our experiments evaluate TemplateCraft's
effectiveness in template generation and reuse and analyze the roles of feedback-driven revision,
long-term memory, and shared persistent assets.

Our main contributions are as follows:
\begingroup
\makeatletter
\appto\@listi{\setlength{\topsep}{3pt}\setlength{\partopsep}{0pt}}
\makeatother
\begin{enumerate}
  \setlength{\itemsep}{1pt}
  \setlength{\parskip}{0pt}
  \setlength{\parsep}{0pt}
  \item We formulate reusable visual-template generation and establish a four-stage pipeline from
  natural-language creative briefs to client-executable templates.
  \item We propose a Planner--Evaluator system that uses execution feedback and memory for
  stage-specific rollback, revision, and cross-task experience reuse.
  \item We construct TemplateBench, a benchmark for reusable visual-template generation, and validate
  the system through comparative experiments, analyzing the roles of iterative revision and shared
  persistent assets.
\end{enumerate}
\endgroup
\par
\endgroup 

\section{TemplateCraft}
\label{sec:templatecraft}

\subsection{Template}
\label{sec:template-definition}

We define a template as a parameterized content program that realizes a particular creative effect.
It composes generative AI tools as needed, accepts user-provided inputs, and produces an image
or video. Formally, a template is represented as
\begin{equation}
T=\langle P,\mathcal{M},W_e\rangle,
\label{eq:template-definition}
\end{equation}
where $P$ is the template specification, which defines the creative effect, input requirements $U$,
and output modality $o\in\{\mathrm{image},\mathrm{video}\}$. The set $\mathcal{M}$ contains the
persistent assets distributed with the template, such as style-reference images or motion-reference
videos. The effect workflow $W_e$ consists of tool invocations and their data dependencies, with
each step recording the tool name, invocation parameters, and output variables. Given a user input
$x\in\mathcal{X}(U)$ that satisfies the input constraints, the instantiated output $y$ is defined as
\begin{equation}
 y=\operatorname{Exec}(W_e;x,\mathcal{M}),\qquad y\in\mathcal{Y}_o,
 \label{eq:template-execution}
\end{equation}
where $\mathcal{X}(U)$ is the permitted input set and $\mathcal{Y}_o$ is the visual output space
associated with modality $o$. During reuse, only the user input $x$ is replaced, while the template
specification, persistent assets, and workflow remain fixed.

\subsection{Template Generation Stages}
\label{sec:template-creation}

We divide template generation into the following four stages.

\noindent\textbf{Template Planning.}
Given a user instruction $I$ and an available tool set $\mathcal{T}$, the planner first retrieves the
category skill most relevant to the request. We construct a template skill library $\mathcal{K}$ by
analyzing the project files and rendered outputs of popular templates in the KwaiCut client. Each
skill summarizes recurring creative concepts, shot organization, the number and types of required inputs, target duration, and
tool combinations for a template category. Template planning is formulated as
\begin{equation}
 P=\operatorname{TempPlan}\bigl(
 I,\mathcal{T};\operatorname{Retrieve}(I,C(I),\mathcal{K})\bigr),
 \label{eq:template-planning}
\end{equation}
where $C(\cdot)$ predicts the category of the user instruction. The retrieval module selects the
corresponding skill from $\mathcal{K}$, and the planner uses it to produce the template
specification $P$.

\noindent\textbf{Material Generation.}
To obtain preview outputs during template generation, the system prepares two types of visual
assets from the template specification $P$. Simulated user inputs can be replaced at
deployment, whereas persistent template assets provide consistent visual or motion conditioning
signals and are stored with the template. The planner reads the asset descriptions in $P$ together with the
available tools $\mathcal{T}$, constructs a material workflow, and delegates its execution to the
executor:
\begin{equation}
 \begin{aligned}
 W_m &= \operatorname{MaterialGen}(P;\mathcal{T}),\\
 \mathcal{S} &= \operatorname{Exec}(W_m).
 \end{aligned}
 \label{eq:material-generation}
\end{equation}
Here, $\operatorname{MaterialGen}(\cdot)$ denotes LLM-based planning for material generation, $W_m$
specifies the preparation steps, and $\mathcal{S}$ is the resulting asset store, which contains the
simulated inputs and persistent assets $\mathcal{M}$, i.e., $\mathcal{M}\subseteq\mathcal{S}$. At this stage, we
require the system to generate persistent assets $\mathcal{M}$ in order to improve template reusability
(see \hyperref[sec:rq3-reference-material]{Experiment 3 (RQ3)}).

\noindent\textbf{Effect Workflow Generation.}
To translate the creative specification into concrete generation operations, the system plans the
effect workflow from $P$, the material store $\mathcal{S}$, and tool descriptions
$\mathcal{T}$. The planner selects tools appropriate for the target effect and specifies the
invocation parameters of each step. The process is formulated as
\begin{equation}
 \begin{array}{@{}l@{\;}c@{\;}l@{}}
 W_e & = & \operatorname{EffectGen}(P,\mathcal{S};\mathcal{T}),\\
 V_{\mathrm{c}} & = & \operatorname{Exec}(W_e;\mathcal{S}).
 \end{array}
 \label{eq:effect-generation}
\end{equation}
Here, $\operatorname{EffectGen}(\cdot)$ uses an LLM to plan $W_e$, and
$V_{\mathrm{c}}$ denotes the resulting preview. The executor resolves variable references to pass
stored assets or preceding outputs to the appropriate tools. Each declared asset must contribute to
the final output in its specified role rather than being used only in an irrelevant intermediate step.

\noindent\textbf{Protocol Compilation.}
To make a template reusable on the client, the system compiles the template
specification $P$, effect workflow $W_e$, and execution-time variable table $\mathcal{R}$ into the
template configuration $\Pi$ after preview execution. The compiler converts variable references into package-local
resource references and translates invocation parameters and step dependencies into a
client-executable operation sequence according to each tool's protocol rules. Compilation and packaging
are expressed as
\begin{equation}
 \begin{aligned}
 \Pi &= \operatorname{ProtoComp}(P,W_e,\mathcal{R}),\\
 \mathcal{B} &= \operatorname{Package}(\Pi,\mathcal{F}).
 \end{aligned}
 \label{eq:protocol-compilation}
\end{equation}
where $\mathcal{F}$ is the set of required asset and audio files, and $\mathcal{B}$ is the final
template package. The compiler distinguishes assets
by their declared roles in $P$: user inputs become replaceable slots, whereas persistent assets such as
reference images and motion videos are retained. The template configuration records the relationship
between these assets and the generation steps and is packaged with all required files.

\subsection{Multi-Agent System}
\label{sec:multi-agent-system}

\begin{figure}[t]
\centering
\includegraphics[width=\columnwidth,trim=19bp 0 28bp 0,clip]{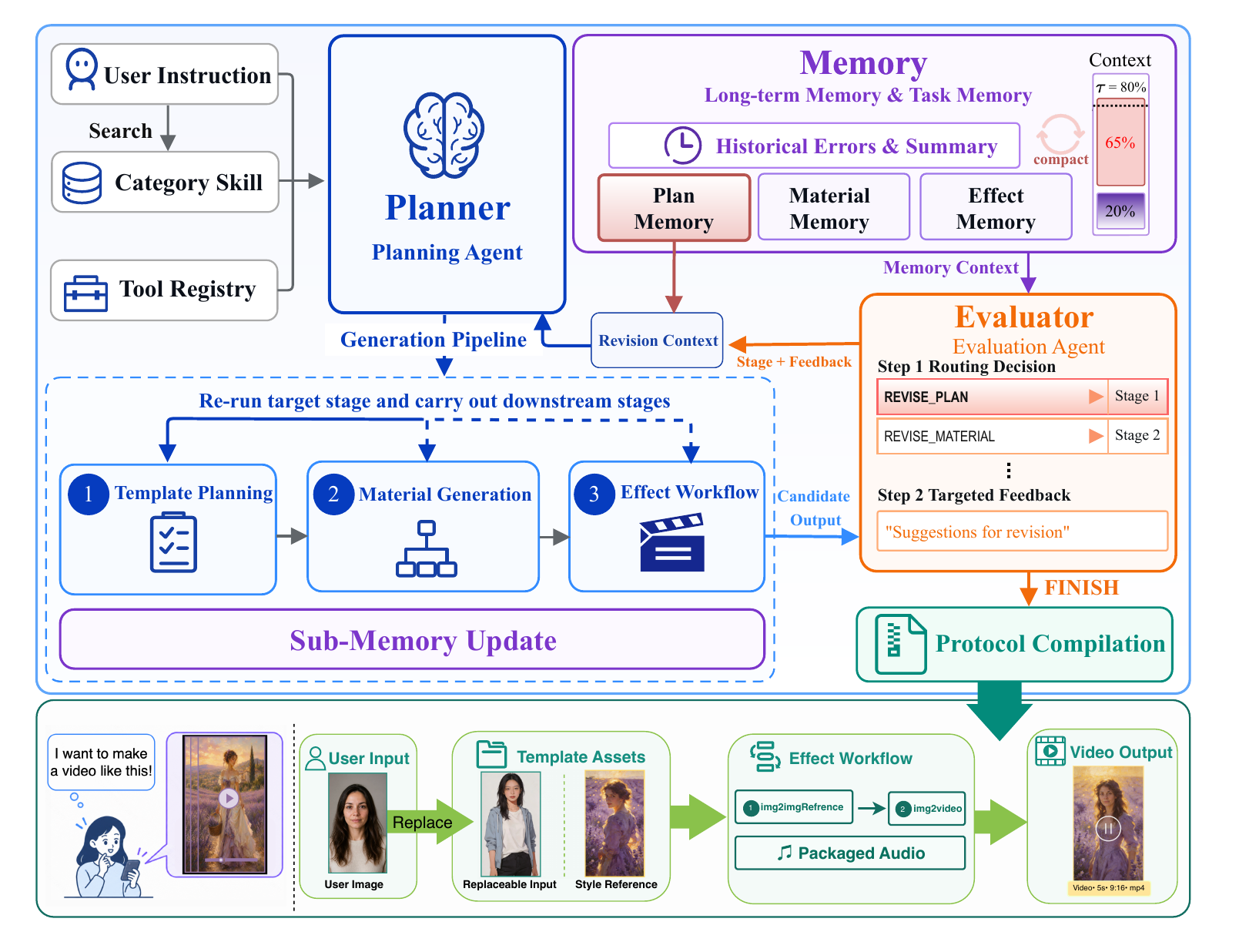}
\caption{The TemplateCraft multi-agent loop for template generation and template reuse.}
\label{fig:agentic-template-creation-loop}
\end{figure}

To improve the reliability of smaller models in multi-stage template generation, TemplateCraft introduces
a Planner--Evaluator loop before protocol compilation. The loop revises specifications and workflows
using execution feedback, while the memory module retains stage states and historical experience, as shown in
Fig.~\ref{fig:agentic-template-creation-loop}.

\subsubsection{Memory}
\label{sec:agent-memory}

Memory comprises task memory and long-term memory. Task memory stores the creative requirements
and records the state of the three generation stages in Plan Memory, Material Memory, and Effect
Memory. After every stage attempt, the system writes the current specification or workflow, execution
result, error information, and stage summary to the corresponding submemory.

Long-term memory is available only to the Evaluator. It records the errors, evaluation feedback,
and terminal states of completed tasks and provides relevant experience for subsequent revisions. When the
Evaluator's input context reaches $\tau$ times the model's context limit, where $\tau\in(0,1)$, the system
consolidates repeated errors and their associated revision traces into a summary that preserves the
principal problems and how they were addressed.

\subsubsection{Planner}
\label{sec:agent-planner}

The Planner generates specifications or workflows according to stage-specific prompts that define
the task, available tools, and output format, and passes the workflows to the executor.
During normal execution, it does not read stage submemory. When a rollback is required, it receives
the target stage selected by the Evaluator together with the associated feedback, retrieves the current artifact and historical summary
from that stage's submemory, and produces a complete revision under the original tool and format constraints.

\subsubsection{Evaluator}
\label{sec:agent-evaluator}

The Evaluator diagnoses problems and determines the revision target from the creative requirements,
execution trace, and stage memories. If execution has produced a candidate output, the Evaluator
performs multimodal assessment using the current trial input. If generation or execution fails before
an output is produced, it localizes the failure based on the current artifacts, completed steps, and error
messages.

Each evaluation round consists of routing and targeted feedback. It first identifies the earliest
stage that must be modified to resolve the problem, then uses that stage's generation prompt, tool
constraints, and long-term memory to provide targeted feedback for revision by the Planner.
The routing action space comprises:
\begingroup
\setlength{\parskip}{0pt}
\makeatletter
\appto\@listi{\setlength{\topsep}{0pt}\setlength{\partopsep}{0pt}}
\makeatother
\begin{itemize}
  \setlength{\itemsep}{0pt}
  \setlength{\parskip}{0pt}
  \setlength{\parsep}{0pt}
\item \texttt{REVISE\_PLAN}:
The overall creative concept, plan structure, or asset design requires adjustment; the system returns to template planning.
\item \texttt{REVISE\_MATERIAL}:
The asset design is sound, but the generation procedure contains errors; the system returns to material generation.
\item \texttt{REVISE\_EFFECT\_WORKFLOW}:
Tool-call or dependency errors require the system to return to effect-workflow generation.
\item \texttt{FINISH}:
A valid candidate template exists and satisfies the creative requirements; the loop terminates.
\end{itemize}
For example, if a persistent asset meets the requirements but is not used in the final generation,
only the effect workflow needs revision; if the asset specification itself violates the creative
requirements, the system returns to template planning.
\par
\endgroup

During rollback, artifacts preceding the target stage are retained. After the Planner completes the
revision, subsequent stages are rerun in order, and the new candidate output or execution error is passed to
the next evaluation round. Evaluation and revision occur only during template generation.

\begin{table*}[!t]
\centering
\caption{Evaluation results of all methods on TemplateBench. Higher is better for all metrics; checkmarks indicate supported generation capabilities, and -- denotes a metric that is not applicable. Bold and underlined values indicate the best and second-best results, respectively.}
\label{tab:main_results}
{\fontsize{9pt}{10.4pt}\selectfont
\setlength{\tabcolsep}{2.5pt}
\renewcommand{\arraystretch}{1.05}
\begin{tabular*}{\textwidth}{@{\extracolsep{\fill}}lcccccccccccc@{}}
\toprule
\multirow[c]{2}{*}{\textbf{Method}}
& \multirow[c]{2}{*}{\shortstack{Video\\Generation}}
& \multirow[c]{2}{*}{\shortstack{Template\\Generation}}
& \multicolumn{5}{c}{Image Template Tasks}
& \multicolumn{5}{c}{Video Template Tasks} \\
\cmidrule(lr){4-8}
\cmidrule(lr){9-13}
& & & $\mathrm{SR}_{c}$ & TA & TR & IQ & AQ & $\mathrm{SR}_{c}$ & TA & TR & IQ & AQ \\
\midrule
Wan2.6
& $\checkmark$ &
& -- & \underline{0.8275} & 0.6411 & 0.7301 & \underline{0.6009}
& -- & 0.4283 & 0.7185 & \textbf{0.7344} & 0.6263 \\
Seed
& $\checkmark$ &
& -- & \textbf{0.8467} & 0.6417 & 0.7289 & \textbf{0.6018}
& -- & 0.5118 & 0.6651 & 0.7201 & 0.6050 \\
Qwen3 + UniVA
& $\checkmark$ &
& -- & 0.7733 & 0.6226 & \textbf{0.7391} & 0.5914
& -- & \textbf{0.7164} & 0.6966 & 0.7184 & 0.6554 \\
Qwen3 + Planner-only
& $\checkmark$ & $\checkmark$
& 56.7\% & 0.5168 & 0.6957 & 0.7186 & 0.5693
& 30.0\% & 0.4034 & 0.7526 & 0.7105 & \underline{0.6574} \\
Qwen3 + Planner-CoT
& $\checkmark$ & $\checkmark$
& 56.7\% & 0.5675 & \textbf{0.7151} & \underline{0.7345} & 0.5636
& \underline{43.3\%} & 0.5481 & 0.7913 & 0.7109 & 0.6544 \\
Qwen3 + Planner-Evaluator
& $\checkmark$ & $\checkmark$
& \underline{63.3\%} & 0.6091 & 0.7010 & 0.7270 & 0.5629
& 36.7\% & 0.6389 & \underline{0.8036} & \underline{0.7259} & 0.6553 \\
Qwen3 + TemplateCraft
& $\checkmark$ & $\checkmark$
& \textbf{66.7\%} & 0.6583 & \underline{0.7021} & 0.7271 & 0.5995
& \textbf{50.0\%} & \underline{0.6534} & \textbf{0.8154} & 0.7159 & \textbf{0.6713} \\
\bottomrule
\end{tabular*}
}
\end{table*}

\section{Experiments}
\label{sec:experiments}

We address three research questions: \textbf{RQ1}: Can generated templates satisfy the creative
requirements and adapt to different user inputs? \textbf{RQ2}: Does TemplateCraft improve template
generation success for a smaller open-weight model? \textbf{RQ3}: Does requiring persistent assets
improve template reusability?

\subsection{TemplateBench}
\label{sec:templatebench}
\label{sec:experimental-setup}

\begin{figure}[t]
\centering
\includegraphics[width=\columnwidth,trim=0 8bp 0 0,clip]{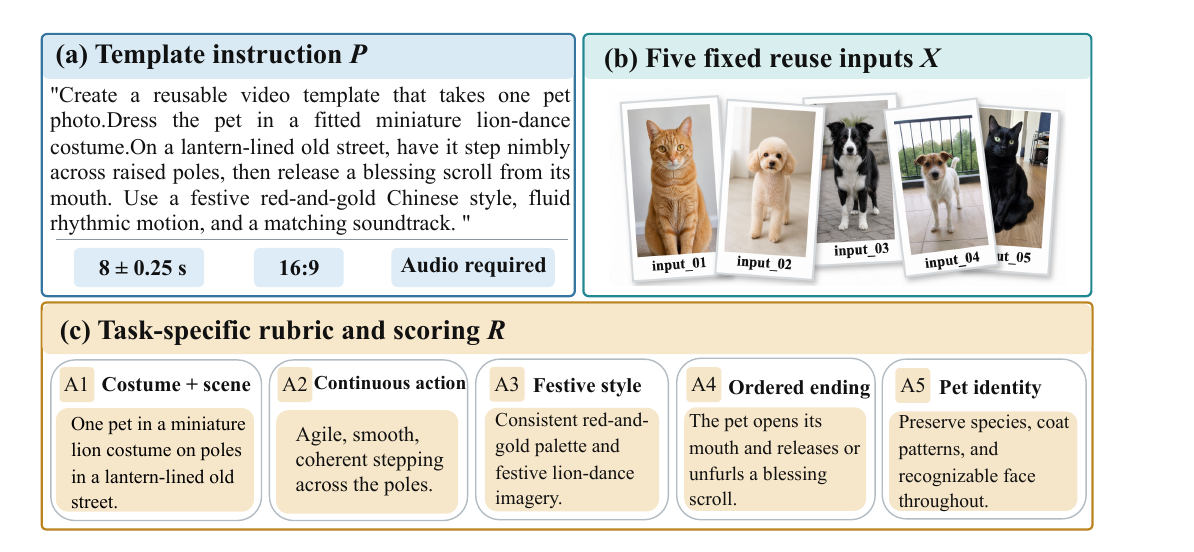}
\caption{An evaluation task with a creative brief, five test-input sets, and a scoring rubric.}
\label{fig:templatebench-overview}
\end{figure}

We use TemplateBench, which comprises 30 image-template tasks and 30 video-template tasks derived
from production templates in KwaiCut. Each task includes a natural-language creative brief, input requirements,
a scoring rubric, and five fixed test-input sets.
No TemplateBench task was used to construct $\mathcal{K}$ or abstract its template skills.
Image tasks cover stylization, composition, and layout; video tasks additionally assess subject interaction,
camera changes, continuity across clips, and audiovisual timing.
Figure~\ref{fig:templatebench-overview} shows an example.

During template generation, the system receives only the textual brief and input requirements and independently
prepares simulated inputs, persistent assets, and a preview. Fixed test inputs remain hidden at this stage.
Once a template is complete, its specification, persistent assets, and workflow are frozen. The template
is then executed on each of the five test-input sets to assess generation reliability and reuse across inputs.

Results are reported separately for each modality. Template generation success rate $\mathrm{SR}_{c}$
measures the proportion of tasks yielding a complete template and a valid preview.
For reuse, we report template adherence (TA), template reusability (TR), image quality (IQ),
and aesthetic quality (AQ). The same Gemini 3.5 Flash evaluator is used for all methods. Given the
creative brief, scoring rubric, test inputs, and final output, it scores task fulfillment and correct input use.
We draw stratified random samples of 60 image outputs and 60 video outputs. Three raters
independently score each output using the same five-point scale, and their mean serves as the
human score. An ordinal Krippendorff's $\alpha$ of 0.743 indicates substantial inter-rater agreement,
while Spearman's $\rho=0.762$ indicates a strong correlation between the VLM and mean human scores.
TR is computed from CSD~\cite{somepalli2024style} descriptors to measure style consistency across outputs of the same
template with different inputs. IQ estimates the perceptual image quality of sampled frames using MUSIQ~\cite{ke2021musiq}
trained on SPAQ~\cite{fang2020spaq}, while AQ measures the visual appeal of sampled frames using the
LAION aesthetic predictor~\cite{laion2022aesthetic}.

\subsection{RQ1: Template Quality}
\label{sec:rq1-template-quality}

We compare five agent methods: TemplateCraft, Planner-only, Planner-CoT,
Planner-Evaluator, and UniVA. Seed and Wan2.6 serve as direct-generation baselines:
\begingroup
\setlength{\parskip}{0pt}
\makeatletter
\appto\@listi{\setlength{\topsep}{0pt}\setlength{\partopsep}{0pt}}
\makeatother
\begin{itemize}
  \setlength{\itemsep}{0pt}
  \setlength{\parskip}{0pt}
  \setlength{\parsep}{0pt}
  \item \textbf{TemplateCraft:} the full system in Section~\ref{sec:templatecraft}, with the memory modules initialized,
  $\tau=80\%$, and at most five revision rounds.
  \item \textbf{Planner-only (best-of-3):} runs the Planner three times without feedback and selects
  the generation-stage preview with the highest TA. We set $N=\lceil 1+1.77\rceil=3$ from one initial run
  and TemplateCraft's average of 1.77 revisions.
  \item \textbf{Planner-CoT (best-of-3):} adds zero-shot CoT prompting~\cite{kojima2022zeroshot}
  to Planner-only, requiring the Planner to analyze tool selection and data dependencies before generation;
  all other settings remain unchanged.
  \item \textbf{Planner-Evaluator:} retains the Planner--Evaluator loop and stage submemories for targeted rollback, but removes long-term memory.
  \item \textbf{UniVA}~\cite{liang2025univa}\textbf{:} a collaborative planning--execution framework
  used as a direct-generation baseline for each creative brief and test-input set.
\end{itemize}
All variants, including TemplateCraft's Planner and Evaluator, use Qwen3-VL-30B-A3B-Instruct~\cite{bai2025qwen3vl}.
Unless otherwise specified, image and video generation use
Seedream 4.0~\cite{seedream2025} and Seedance 1.5 Pro~\cite{seedance2025}, respectively. In addition to the Seed model family, we include the
Wan2.6 family as a direct-generation baseline. For video-template tasks with multiple inputs, each
direct-generation baseline uses the corresponding multi-image-reference model, namely Seedance
2.0~\cite{seedance2026} or Wan2.6-R2V-Flash, under the same evaluation protocol.
\par
\endgroup

In Table~\ref{tab:main_results}, TemplateCraft leads template-generating methods in TA, improving
image/video TA from 0.5168/0.4034 to 0.6583/0.6534 over Planner-only. Its TR gains also indicate
better reuse. Planner-CoT has higher image TR but lower TA, showing that CoT cannot replace
execution-driven revision. Seed and UniVA each achieve higher TA in one modality but do not generate
reusable templates. Perceptual-quality leaders vary by metric: UniVA, Seed, Wan2.6, and TemplateCraft
lead image IQ, image AQ, video IQ, and video AQ, respectively. TemplateCraft's main gains therefore
lie in template-generation reliability, adherence, and reuse rather than generator-dependent perceptual quality.

\subsection{RQ2: Multi-Agent Capabilities}
\label{sec:rq2-multi-agent}

We assess whether TemplateCraft improves template generation with a smaller open-weight model through
ablations and comparison with a GPT-4o~\cite{openai2024gpt4o} Planner-only baseline, using the
settings in Section~\ref{sec:rq1-template-quality}.

\noindent\textbf{Ablation experiments.}
As shown in the left panel of Fig.~\ref{fig:template-performance-csd}, Planner-CoT leaves image
$\mathrm{SR}_{c}$ unchanged but raises video $\mathrm{SR}_{c}$ from 30.0\% to 43.3\%, suggesting
that explicit reasoning benefits complex video workflows. Planner--Evaluator raises image
$\mathrm{SR}_{c}$ to 63.3\%, indicating that targeted rollback corrects some execution errors. With
long-term memory, TemplateCraft reaches 66.7\%/50.0\% on image/video tasks, gains of 10.0/20.0
percentage points over Planner-only, and achieves the highest TA among the Qwen-based template-generation variants. These results
show that execution feedback, stage rollback, and historical experience jointly improve template
generation success, with larger benefits on more complex video workflows.

\noindent\textbf{Model comparison.}
Qwen-based TemplateCraft matches GPT-4o Planner-only in image $\mathrm{SR}_{c}$ and improves video
$\mathrm{SR}_{c}$ from 40.0\% to 50.0\% and TA from 0.6383 to 0.6534, although its image TA is lower.
It thus narrows or reverses the gap on selected metrics, with additional evaluation and revision calls.

\subsection{RQ3: Persistent Assets}
\label{sec:rq3-reference-material}

We use all 60 template tasks in TemplateBench to compare two settings: explicitly requiring the
system to generate and use persistent assets, and imposing no such requirement, in which case the
model decides whether to use persistent assets. We measure cross-input style consistency using TR.

\begin{figure}[!t]
\centering
\includegraphics[width=\columnwidth]{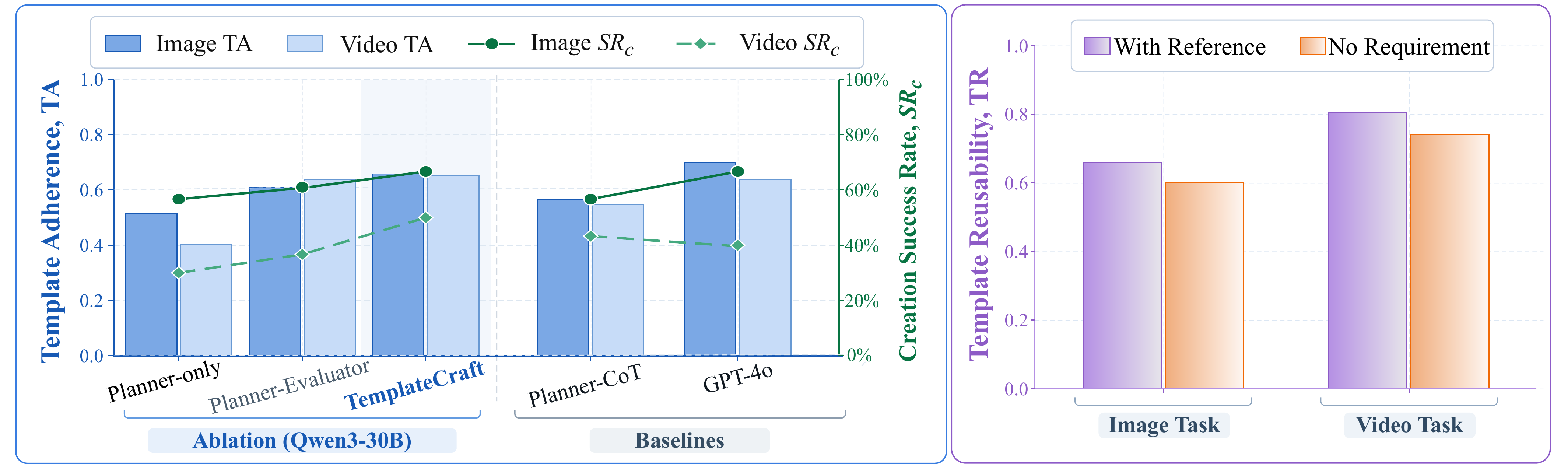}
\caption{TemplateCraft ablation and capability comparison, together with the persistent-asset experiment. Left: template
adherence (TA) and template generation success rate ($\mathrm{SR}_{c}$) for different
system configurations and the GPT-4o baseline on image and video tasks. Right: template reusability (TR)
with and without an explicit requirement to use persistent assets.}
\label{fig:template-performance-csd}
\end{figure}

As shown in the right panel of Fig.~\ref{fig:template-performance-csd}, the explicit requirement improves TR for both modalities,
with a larger gain on video tasks. Shared persistent assets therefore help preserve output style
across different inputs.

\section{Conclusion}

We introduced TemplateCraft, which supports reusable visual-template generation through a staged
pipeline, feedback-driven revision, and long-term memory. With the same Qwen3-VL backbone,
TemplateCraft improves the template generation success rate and adherence over Planner-only for both image and video
templates, and achieves the highest adherence and video reusability among template-generating methods.
Ablations suggest that targeted revision from execution feedback and long-term memory jointly improve
the smaller open-weight model's template-generation capability. The full Qwen-based system matches or exceeds
GPT-4o with Planner-only on selected metrics, although this comparison includes additional evaluation
and revision calls. The persistent-asset experiment further shows that shared persistent assets help
preserve style across inputs. These results show that TemplateCraft can convert natural-language
creative instructions into reusable visual templates through a multi-agent system.

\begingroup
\fontsize{9pt}{10.4pt}\selectfont
\section{Compliance with Ethical Standards}

This study involves no biomedical intervention or animal experimentation. To validate VLM-based
template-adherence scores, three raters independently assessed a stratified sample of 120 TemplateBench
outputs using a shared five-point rubric, with method identities and implementations concealed. Only task
scores were recorded, and only aggregate agreement is reported. TemplateBench contains abstract specifications
and rubrics derived from KwaiCut usage patterns, but no original client assets or user-uploaded content.
Test assets and generated outputs were used under applicable permissions; TemplateCraft is not intended
for high-stakes decisions about individuals.

\section{Acknowledgments}

This work was funded by Kuaishou Technology. Several authors are affiliated with Kuaishou Technology.
The authors declare no other competing financial interests.

\bibliographystyle{IEEEbib}
\bibliography{strings,refs}
\endgroup
\end{document}